# Computerized Multi Microphone Test System

A. M. Dorman, test engineer, consulting, Israel

**Abstract**. An acoustic testing approach based on the concept of a microphone sensor surrounding the product under test is proposed. Partial microphone signals are processed simultaneously by a test system computer, according to the objective of the test. The spatial and frequency domain selectivity features of this method are examined. Sound-spatial visualization algorithm is observed. A test system design based on the concept of a microphone surrounding the tested product has the potential to improve distortion measurement accuracy in a noisy ambience, to meet spatial resolution requirements for acoustic inspection.

**Introduction.** Microphone arrays are used for acoustic wave source position localization (e.g. in seismology, hydro-acoustic systems). The spatial selectivity feature of acoustic array systems may help to suppress acoustic noise. This feature of microphone arrays finds its application in mobile voice communication. This article focuses on design issues of computerized testers used for automated testing of acoustic characteristic of production (e. g. in acoustic test of mobile communication devices like cell phones). The **P**roduct **U**nder **T**est (**PUT**), which is mounted in a tester, should be configured and activated to output acoustic waves. In conventional testing, for example, a harmonic signal is applied to the **PUT** speaker, while the tester microphone transducer picks up signals produced by the acoustic wave. The electrical output signal of the microphone is analyzed to verify whether its parameters, as sound pressure level and distortion, are within specified limits. Accurate acoustic test measurements, especially distortion, (e.g. **T**otal **H**armonic **D**istortion (**THD**), should be performed in an acoustic chamber where extraneous acoustic waves are thoroughly suppressed on the fundamental frequency and its harmonics, as well. The weakness in the traditional testing concept is a lack of robustness, especially if measurements are carried out without sufficient acoustic isolation from ambient noise.

**Concept.** The acoustic testing concept considered in this article is based on employment of a microphone array. In accordance with the concept proposed here, the microphone array surrounds **PUT**; unlike in the classical concept, **PUT** is enclosed by the microphone sensor. This kind of sensor is termed here the Enclosing Microphone (**EM**). **EM** may be implemented practically, not only as an array of microphones, but also as a net structure membrane.

The **EM** technique permits potential reduction of external acoustic noise, improved quality of acoustic product testing, and improved test robustness, without significant acoustic isolation. It has also some other useful features, such as spatial resolution. Spatial resolution is an inherent feature of the **EM** technique. A test system equipped with **EM** may provide additional diagnostic capabilities (e.g., monitoring of production quality, possible enhancement of **PUT** assembly failure detection) and more. This article is based on the author's unpublished papers from the period 1992 to 2003.

**Analysis**. It is assumed that a microphone array or net does not significantly deform acoustic waves. Figure 1 shows a simple **EM** configuration. Figure 1 comprises eight identical **EM** partial media pressure microphones, on a cube with vertexes **M1... M8**.

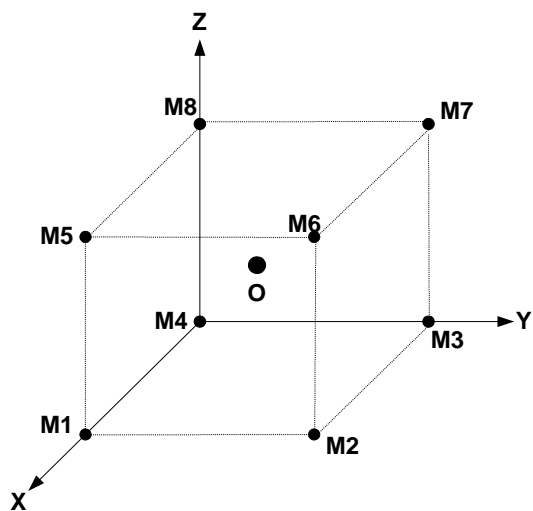

**Figure 1. Simple EM configuration**

**PUT** is in the center **O** of this cubic **EM**. The microphones have the same isotropic directional diagrams and sensitivities. The Cartesian **X**, **Y** and **Z** axes as represented in Figure 1 are directed as so that microphone **M4** is at the origin and microphones **M1**, **M3**, **M8** lie on **X**, **Y** and **Z** axes, respectively. Figure 2 is a diagram of the tester.

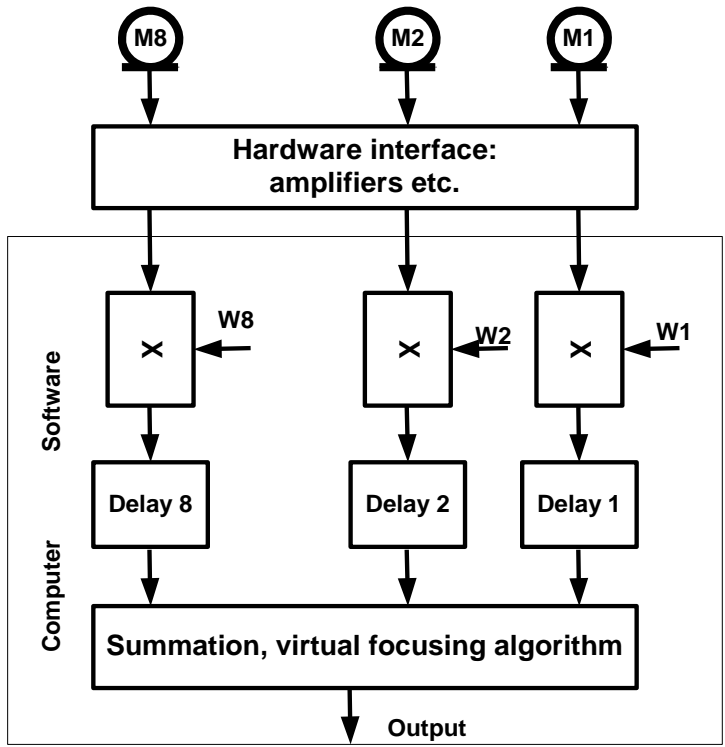

**Figure 2. Tester block diagram**

Each **EM**-output signal is delayed and multiplied by weights **W1** ... **W8**, and added to the **EM** tester total (each microphone, **M1** … ,**M8,** is a partial sensor). This total output signal is used to test **PUT** frequency response, distortion etc. Let us suppose that an isotropic spherical wave is radiated from **PUT** location **O,** that all weights and delays are equal, and as mentioned above, that all microphones (signals from each of which are part of the **EM** total) are sufficiently small that none of them distorts the acoustic wave. Therefore, the total **EM** tester output signal is eight times that of the classical single microphone sensor. A tester incorporating such **EM** has flat frequency response, assuming that partial microphone frequency responses are equalized. Let us assume that **EM** is exposed to an external plane acoustic wave, and that its propagation direction is parallel to one of the axes (**X, Y , Z**).

If cube of edge length **d** is equal to half of wave length $\lambda$, then the tester output signal is zero. The signals on the odd harmonics of this frequency will also be rejected. The wave lengths $\lambda$ of these signals may be found from the following equation,

$$0.5n\lambda = d,$$

where **n** = 1, 3, 5... .

Therefore desirable frequencies **f** for accurate acoustic measurement and testing with noisy outside ambience may be given in the form

$$f = c/\lambda = cn/2d, \qquad (1)$$

where **c** is sonic speed **[1]**.

If fundamental is **c/2d** and the tester output signal is filtered by a narrowband hardware or software filter on frequencies given by ( **1** ), distortion may be accurately measured in a nosy ambience, even without acoustic isolation (various **EM** edge length should be used for even harmonic amplitude measurements: **d/2** for $2^{nd}$ harmonic, etc.).

Such a cubic **EM** also rejects external plane waves with other directions of propagation. Therefore, **EM** serves as a notch (rejecting) filter for external acoustic waves. But the rejection frequency depends on the wave propagation direction. If, for example, the plane wave front is parallel to the plane containing microphones **M1**, **M3**, **M7** and **M5** (Figure 1), the signal rejection condition would be

$$f = cn/2^{0.5}d.$$

For maximum suppression of external ambient acoustic waves, **EM** cube orientation and its edge length **d** should be optimized in accordance with the spatial spectrum of outside radiation. Every sound source radiates complicated waves, which may be represented as the superposition of plane or spherical harmonic waves. Therefore, the **EM** transfer function, for harmonic waves, plane or spherical, should be studied.

**Spherical sensor**. Cubic **EM** is a special case of spherical type where all **EM** microphones are arranged in a sphere around **PUT**. The number of **EM** microphones on a sphere should be increased for the purpose of yielding rejection of ambient external waves for different directions of propagation. Assume that the number of **EM** microphones may be increased without limit. This approach is useful for estimating possible features of the **EM** method, and may serve as a practical design guide with a limited number of **EM** microphones. Figure 3 depicts **EM** as a spherical capacitance microphone.

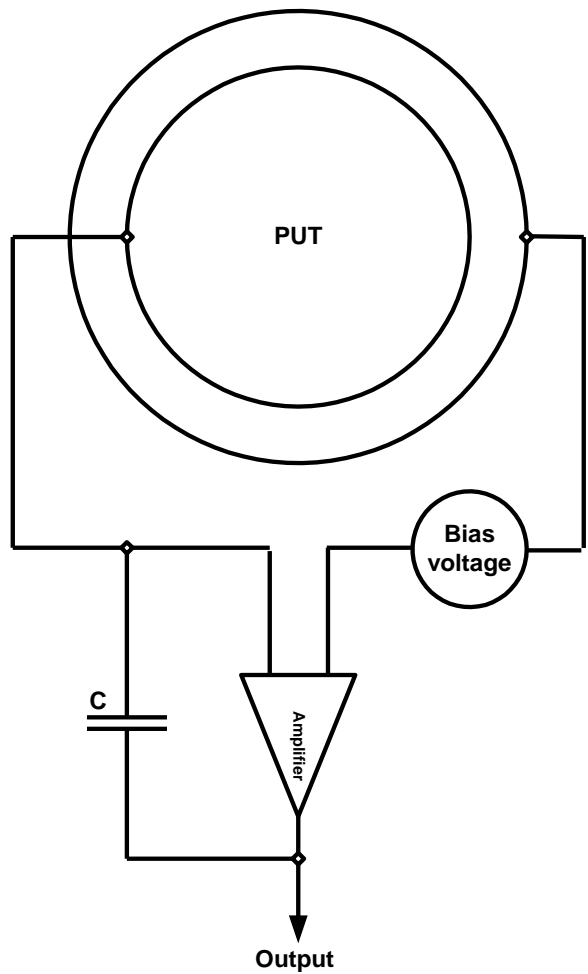

**Figure 3. Spherical capacitance microphone enclosing PUT**

**PUT** position in Figure 3 is in **EM** sphere center **O**. The internal electrode is biased by direct current voltage, and the external electrode is grounded. **PUT** acoustic wave pressure modulates capacitance between electrodes, so that **EM** converts acoustic pressure into an electric signal.

**Frequency domain features.** Assume that **EM** is exposed to an external plane harmonic wave and all **EM** microphone channels have equal delays and weights. The **EM** output signal is a sum of weighted and delayed output signals of all **EM** microphones. Suppose, as earlier, that **EM** microphones are sufficiently small for **EM** transparency and do not distort external waves. This means that with condenser **EM** (Figure 3), both internal and external electrodes are transparent to acoustic waves. This idealization may help to evaluate the principle nature of **EM**. Taking into account this idealization, it can be proven that the resulting **EM** (tester) output signal is

$$I_h(t) = A_h \cos\Phi_0(t), \qquad (2)$$

where amplitude $A_h = S_0 P_0 R(f)$,

is a product of sound pressure $P_0$ of external wave, integral sensitivity $S_0$ of **EM**, and the transfer function $R(f)$ for the external plane wave.

$$R(f) = \sin(2\pi f r/c)/(2\pi f r/c),$$

where **r** is the radius of the **EM** sphere.

$\Phi_0(t)$ is a phase value of the external acoustic wave in **EM** center **O**, assuming no obstacles to its propagation.

It may be shown that this result is also applicable to external spherical harmonic waves. The transfer function is independent of the direction of wave propagation. Now suppose that **PUT** is placed at the **EM** center, and that it radiates an isotropic spherical wave. Tester response to this acoustic wave stimuli is

$$I_{put}(t) = A_{put} \cos[\Phi_{put}(t) - 2\pi f\, r/c],$$

with amplitude $A_{put} = S_0 P_{put}$.

$2\pi f r/c$ is a phase shift for the wave propagation distance **r** from the center **O** of the **EM** sphere to its surface. $P_{put}$ is a sound pressure of the **PUT** wave on the **EM** surface. $\Phi_{put}(t)$ is a phase of the **PUT** wave in **O**.

These equations help find the noise to signal ratio **N/S**, where noise is an external interfering wave and signal is a sound wave generated by **PUT**. Noise to signal ratio **N/S** is

$$N/S = (A_h / A_{put}) R(f).$$

The $A_h / A_{put}$ factor in this equation is proportional to the sound pressure ratio $P_0 / P_{put}$ of an interfering external wave and a **PUT** generated wave. The second factor is the transfer function **R(f)** for an external interfering wave.

The plot of this function is shown in Figure 4, where the abscissa is $X = 2\pi f r / c$.

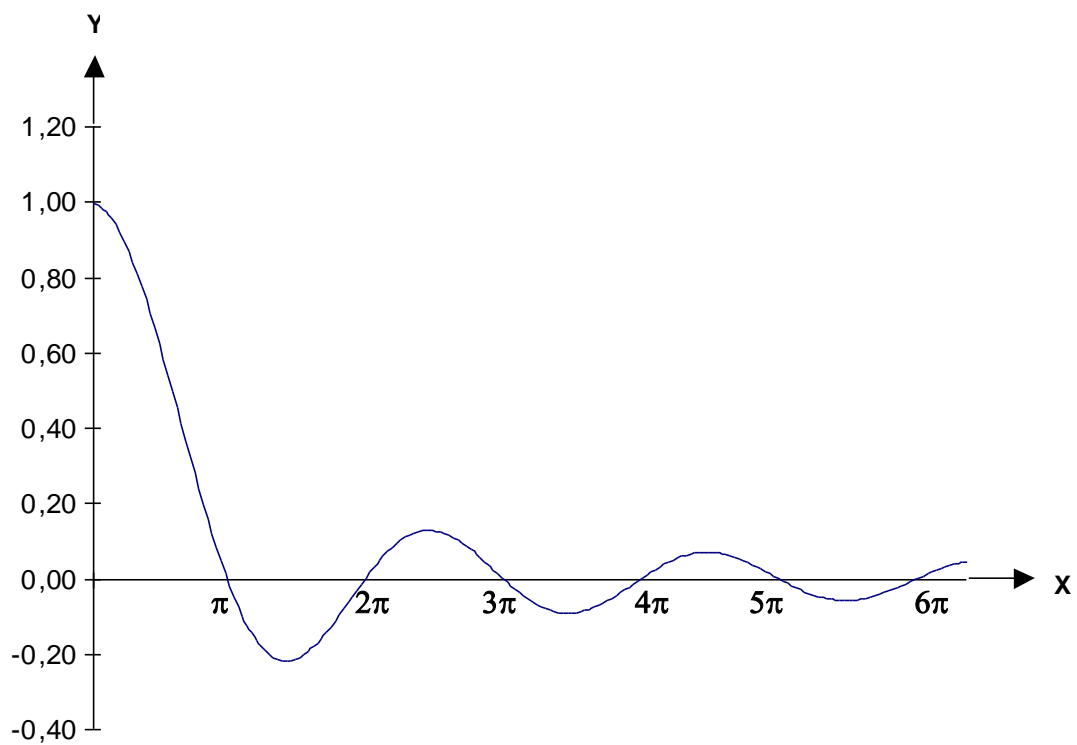

**Figure 4. EM transfer function (Y=R(f)) for external wave.
X is proportional to the frequency f .
Alternatively,
EM spatial resolution function (Y=W(e₀)).
X is proportional to the distance e₀ from the EM sphere center**

Suppose that we select fundamental frequency $f_0$ for **PUT** testing, thus:

$$f_0 = c/2r. \qquad (3)$$

For this frequency, **R(f) = 0** (for fundamental and its harmonics). This results in dramatic improvement of noise to signal ratio and therefore in the accuracy of testing: **THD** and frequency response tests. If for example **r** = 10 cm, the fundamental frequency choice should be about 1.7 kHz.

Similar results may be obtained for **EM** which screens an external interfering wave. Assume that the **EM** sphere, "illuminated" by an external plane wave, obscures the other shaded **EM** hemisphere. In this case, the transfer function for the external wave is

$$R(f) = \sin(\pi f r/c)/(\pi f r/c) .$$

The first zero transfer function frequency will be twice that given by ( **3** ). It should be emphasized that ( **3** ) is similar to ( **1** ), given a simple cubic **EM**. The equation is more complicated for the transfer function when less than a hemisphere of **EM** is illuminated by an external wave.

Suppose that only a portion of an **EM** sphere surface is sensitive and contributes to the output signal. This area is defined by the intersection of the plane and **EM** sphere, and observed from the location of external spheroid wave source. The plane is normal to the direction, from **EM** center, to the external spherical wave source. The **EM** output signal transfer function in this case is

$$R(f) = const * \sin[((\pi f R/c)(1+\alpha^2 - 2\alpha \cos(\varphi_0))^{0.5} - 1 + \alpha]/(\pi f r/c),$$

where **R** is the distance between the external wave source and the center of **EM**,

**α=r/R**,

**$\varphi_0$** is the angle between the rim of the sensitive **EM** area and the direction from the **EM** center to the external wave source.

It should be noted that the area of an **EM** sphere which contributes microphone signals to an **EM** output signal may be controlled by a tester computer, to artificially simulate **EM** shading ; as a result, it can accurately very rejection frequency.

**Spatial resolution**. An inherent feature of **EM** is spatial resolution, which should be analyzed for several reasons, among them: to understand and specify test station requirements, and to estimate the accuracy of sound pressure measurements, especially when the geometric dimensions of **EM** resolution and **PUT** are approximately the same. In particular, it is important to understand whether or not **PUT** can be considered a spot source radiating spheroid acoustic waves. According to Huygens' principal, an acoustic wave radiated by **PUT** is a superposition of several spheroid waves, whose virtual sources are distributed, for example on the surface of **PUT**. The **EM** response in this case depends on its spatial resolution properties. **EM** is a linear system. The **EM** response in such complex cases is the sum of the responses to partial spheroid waves, which represent wave stimuli radiated by **PUT**. The **EM** response to a trial spheroid acoustic wave the spot source of which is located inside the **EM** and normalized by its maximum value, is an **EM** resolution function. Resolution function argument $e_0$ is the bias, or shift, of the source from the geometric center **O** of **EM**. The maximum response value is when bias is zero, i.e., when the source is the center of **EM**. It may be shown that the spatial resolution function, as a function of a spot source shift $e_0$ of a spheroid trail wave from the **EM** center, is given by the equation

$$W(e_0) = \sin(2\pi f\, e_0/c)/(2\pi f\, e_0/c). \qquad (4)$$

The spatial resolution function (the ordinate) is shown in Figure 4, where the abscissa is equal to $2\pi f\, e_0/c$. Geometric center **O** of **EM**, where the absolute response value is at a maximum, may be termed the **EM** focus. The response function is a real-valued function. The tester output signal phase depends on bias $e_0$ as follows. The resolution function phase changes in value from **0** to $\pi$ (or from $\pi$ to **0**) at points where the absolute value of resolution function $|W(e_0)|$ is a zero (solution).

It follows from **(4)** that tester output signal amplitude reduces to zero if the frequency of the trial acoustic signal is

$$f = n \ (c/2e_0). \qquad (5)$$

The radius of a sphere $e_0$, which may be found from this equation for given frequency **f**, may be termed the spatial resolution radius of **EM**. It is the radius of a first spatial resolution function zero. The spatial resolution radius does not depend on the radius of **EM**, but rather on the testing frequency **f**. If, for example, test frequency **f** is 10 kHz, the spatial resolution radius is approximately 3.4 cm. This equation is useful for estimating whether **PUT** is sufficiently small to be considered a spot radiator, or should be considered one more complex .

**Acoustic virtual imaging.** **EM** spatial resolution features my be used for non-intrusive **PUT** acoustic testing. For example, such inspection may be for **PUT** assembly failures, including **PUT** case cracks. The result is spatial change in **PUT** acoustic radiation. Spatial scanning can be used to obtain an acoustic image. This involves the **EM** focus scanning over the **PUT** corpus. The main contribution to the **EM** output signal comes from the **PUT** area being in the **EM** focus. The radius of this area is approximately $e_0$ , as given in ( **5** ).

**PUT** body acoustic scanning may be implemented by mechanical shift of **PUT** or **EM**. This test method assumes data acquisition for all **EM** focus positions relative to **PUT**.

Another test method, termed **virtual scanning**, requires less time. Assume that all **EM** partial microphone output signals are acquired separately and simultaneously via hardware interface, for example as represented in Figure 1; they are saved in computer memory. The hardware interface may also include multiplexer (not shown). These acquired data may be termed the acoustic record. Only one acoustic record saved in computer memory is required for virtual **EM** focus scanning. This single record is used to reconstruct the **PUT** acoustic image. There is no **EM** mechanical scanning with corresponding signal acquisition for all **EM** focus positions. This results in test time reduction. Moreover tests are less susceptible to instrument and **PUT** parameters drifts.

Figure 5 represents a spherical **EM** intersected by plane **AA**, passing through the points **O** and **O'**. Point **O** is the **EM** center. Point **A** is on the sphere of **EM**.

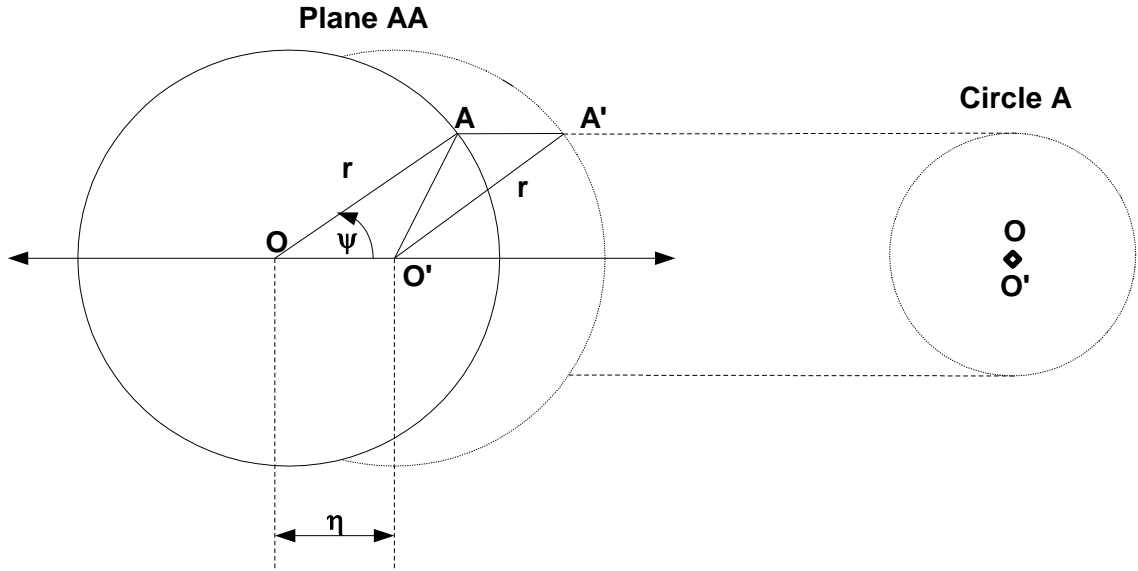

**Figure 5. Virtual focusing algorithm**

The partial **EM** microphone at point **A** receives signal $S_A(t)$, which is kept in computer memory. Assume that **EM** must be refocused on point **O'**. This is possible by mechanical moving **EM** into a new focus position at point **O'**.

If the **EM** center scans mechanically from initial point **O** to point **O'**, **EM** microphone **A** shifts at point **A'**, as shown in Figure 5.

The refocusing of a sensor at any point without its mechanical shift may be named virtual focusing.

In order to virtually set the **EM** focus to **O'**, microphone **A** output signal should be delayed by time $\tau_A$ and amplified by gain $K_A$, as if this signal was received from partial microphone **A'** of **EM**, the center of which has been shifted mechanically to **O'**.

According to Figure 5

$$\tau_A = (|A'O'| - |AO'|)/c + \tau_0.$$

Here $\tau_0$ must be chosen so as to assure a positive $\tau_A$ value for all partial microphones on sphere **EM**. $AO = A'O' = r$, where **r** is the **EM** sphere radius. The same delay $\tau_A$ and gain $K_A$ should be applied to all microphones **A**, which may be seen from **EM** center **O** at angle of $\psi$, and lies on a circle **A** traced by a point **A** when plane **AA** rotates around the axis **OO'** (Figure 5). If distance **OO'** from **EM** center **O** to virtual tester focus **O'** is $\eta$, delay and gain according to Figure 5 are as follows

$$\tau_A = \tau(\psi, \eta) = [r - (r^2 + \eta^2 - 2r\eta\cos\psi)^{1/2}/c] + \tau_0,$$

$$K_A = K(\psi, \eta) = r/(r^2 + \eta^2 - 2r\eta\cos\psi)^{1/2}.$$

The delay and gain should be applied to all signals acquired by partial microphones on a circle **A**. After this transformation, these signals are summed thus

$$v_A(t) = \sum K_A S_A(t - \tau_A).$$

The additive effect

$$v(t) = \sum v_A(t),$$

calculated for all circles **A** of sphere **EM** ($\psi \in [0,\pi]$), represents the **EM** response when **EM** is virtually focused on point **O'**. The shape of the spatial resolution function obtained by virtual scanning differs from that obtained from equation ( **4** ). The greater the $\eta/r$ ratio, the more the virtually focused **EM** spatial resolution function differs from that given by ( **4** ).

**Conclusion**

**1**. **EM** suppresses external acoustic hindrances on specific frequencies. This feature leads to a reduction of the acoustic reflection impact on testing quality. **EM** sensing leads to improved measurement accuracy, especially **THD**. **EM**-based test methods may be useful for acoustic visualization and non-intrusive acoustic inspection.

**2**. The **EM** method and related equations may help find a compromise between requirements for test accuracy and robustness, on the one hand, and tester complexity (e.g. number of **EM** partial microphones), on the other.

**3**. Future development of the **EM** method may include the creation of volumetric sensors, designed as a set of nested spherical (or other-shaped) **EM** sensors, with not-uniform or arbitrary positioning of partial microphones. This concept would permit an improved shape of the major lobe of the **EM** spatial resolution function and suppression of the sidelobes.

**References**

**1**. Горелик Г. С. , Колебания и волны , Москва, 1959